\documentstyle[11pt,aasms4,rotating]{article}

\def\simlt{\lower.5ex\hbox{$\; \buildrel < \over \sim \;$}}
\def\simgt{\lower.5ex\hbox{$\; \buildrel > \over \sim \;$}}

\def\gcm3{{\rm\,g\,cm^{-3}}}
\def\ncm3{{\rm\,cm^{-3}}}

\def\>{$>$}
\def\<{$<$}

\lefthead{Melia et al.}
\righthead{}

\begin{document}
\centerline{Submitted to the Editor of the Astrophysical Journal Letters}
\vskip 0.5in
\title{\bf Spin-Induced Disk Precession in the Supermassive\\
Black Hole at the Galactic Center}

\author{Siming Liu\altaffilmark{1} and Fulvio Melia$^{1,2,3}$}

\affil{$^1$Physics Department, The University of Arizona, Tucson, AZ 85721}   
\affil{$^2$Steward Observatory, The University of Arizona, Tucson, AZ 85721}

% Notice that each of these authors has alternate affiliations, which
% are identified by the \altaffilmark after each name.  The actual alternate
% affiliation information is typeset in footnotes at the bottom of the
% first page, and the text itself is specified in \altaffiltext commands.
% There is a separate \altaffiltext for each alternate affiliation
% indicated above.

\altaffiltext{3}{Sir Thomas Lyle Fellow and Miegunyah Fellow.}

% The abstract environment prints out the receipt and acceptance dates
% if they are relevant for the journal style.  For the aasms style, they
% will print out as horizontal rules for the editorial staff to type
% on, so long as the author does not include \received and \accepted
% commands.  This should not be done, since \received and \accepted dates
% are not known to the author.

\begin{abstract}

Sgr A* is a compact radio source at the Galactic Center, thought to be the
radiative manifestation of a $2.6\times 10^6\; M_\odot$ supermassive black
hole. At least a portion of its spectrum---notably the mm/sub-mm
``bump''---appears to be produced within the inner portion ($r< 10\,r_S$)
of a hot, magnetized Keplerian flow, whose characteristics are also
consistent with the $\sim 10\%$ linear polarization detected from this
source at mm wavelengths. (The Schwarzschild radius, $r_S$, for an object
of this mass $M$ is $2GM/c^2\approx 7.7\times 10^{11}$ cm, or roughly
$1/20$ A.U.) The recent detection of a 106-day cycle in Sgr A*'s radio
variability adds significant intrigue to this picture, since it may signal
a precession of the disk induced by the spin $a$ of the black hole.  The
dynamical time scale near the marginally stable orbit around an object
with this mass is $\approx 20$ mins. Thus, since the physical conditions
associated with the disk around Sgr A* imply rigid-body rotation, a
precession period of 106 days may be indicative of a small black hole spin
if the circularized flow is confined to a region $\sim 30\,r_S$, for which
$a\approx (M/10)\ (r_o/30\ r_S)^{5/2}$.  The precession of a larger
structure would require a bigger black hole spin. We note that a small
value of $a/M$ ($< 0.1$) would be favored if the non-thermal ($\sim 1-20$
cm) portion of Sgr A*'s spectrum is powered with energy extracted via a
Blandford-Znajek type of process, for which the observed luminosity would
correspond to an outer disk radius $r_o\sim 30\ r_S$.  Such a small disk
size is also suggested by earlier hydrodynamical simulations, and is
implied by Sgr A*'s spectral and polarimetric characteristics.

\end{abstract}

% The different journals have different requirements for keywords.  The
% keywords.apj file, found on aas.org in the pubs/aastex-misc directory,
% contains a list of keywords used with the ApJ and Letters.  These are
% usually assigned by the editor, but authors may include them in their
% manuscripts if they wish.


% That's it for the front matter.  On to the main body of the paper.
% We'll only put in tutorial remarks at the beginning of each section
% so you can see entire sections together.

% In the first two sections, you should notice the use of the LaTeX \cite
% command to identify citations.  The citations are tied to the
% reference list via symbolic KEYs.  We have chosen the first three
% characters of the first author's name plus the last two numeral of the
% year of publication.  The corresponding reference has a \bibitem
% command in the reference list below.
%
% Please see the AASTeX manual for a more complete discussion on how to make
% \cite-\bibitem work for you.

\section{Introduction}

Sgr A*'s radio emission is known to be variable, with a fluctuation
amplitude that increases toward high frequencies (Zhao et al. 1993). At
mm wavelengths, large-amplitude ($\sim 100\%$) variations have also been
observed (Wright \& Backer 1993; Miyazaki et al. 1999). Its time-averaged
spectrum is roughly a power law below $100$ GHz, with a flux density
$S_\nu\propto\nu^\alpha$, and $\alpha\sim 0.19-0.34$. In the mm/sub-mm
region, however, Sgr A*'s radiative output is dominated by a ``bump'' 
extending above this power law (Zylka et al. 1992; Falcke et al. 1998).

Unlike the high level of polarization seen in this object at mm/sub-mm
wavelengths (Aitken et al. 2000), Sgr A* reveals a lack ($<1\%$) of
linear polarization below $112$ GHz, though some circular polarization
($\sim 1\%$) has been detected (Bower et al. 1999; Bower et al. 2001).
These prominent spectral and polarimetric differences (Melia, Bromley,
\& Liu 2001) between the cm and the mm/sub-mm bands suggest two different 
emission components in Sgr A*. Because higher frequencies correspond to 
smaller spatial scales (Melia 1992; Melia et al. 1992; Narayan et al. 1995), the 
mm/sub-mm radiation is likely produced in the vicinity of the black hole. 
Earlier work (e.g., Melia 1992, 1994; Coker \& Melia 1997) has indicated that Sgr A* 
is accreting from the stellar winds surrounding the black hole and that the 
infalling gas circularizes at a radius of $\sim 20-800\ r_S$. We recently 
(Melia, Liu, \& Coker 2001; Bromley et al. 2001) demonstrated that the inner 
$10\,r_S$ of the resultant Keplerian structure can not only account for 
the mm/sub-mm properties via thermal synchrotron emission, but it may 
also produce Sgr A*'s X-ray spectrum in the quiescent state (Baganoff et 
al. 2001) via Comptonization of the mm/sub-mm photons. On the other hand, 
the cm radio emission appears to be produced by non-thermal synchrotron 
emission (Liu \& Melia 2001).  Meanwhile, the recently detected X-ray 
flare (Baganoff et al. 2001) from Sgr A* appears to be produced by 
thermal bremsstrahlung radiation resulting from a transient injection of
additional mass flowing through the disk (Liu \& Melia 2002).

The 106-day cycle seen at $1.3$ and $2.0$ cm (Zhao et al. 2001) is
intriguing because high resolution VLA observations have already ruled out
the possibility that such a period might be produced by an orbiting
emitting object, which would have been distinguished spatially from the
main source (Bower et al. 1998).  This inference is supported by the
observed lack of proper motion (Backer et al. 1999; Reid et al. 1999) of
Sgr A*, which precludes its possible association with rapidly moving
components. In addition, any stellar origin for such a source falls short
of the power required to account for the measured radio emission.  It is
reasonable to conclude, therefore, that the periodic variations in Sgr A*
are intrinsic to the source.  In this {\it Letter}, we examine the
possibility that the observed 106-day radio modulation is associated with
a precession of the disk in a Kerr metric and explore the implications
this may have on the nature of Sgr A*.

\section{Nature of the Long-Term Modulation}

The characteristics of this 106-day cycle constrain the nature of its
origin rather tightly (Zhao et al. 2001). First, the observed period is
independent of the wavelength, and recent Submillimeter Array (SMA) and
VLA observations of Sgr A* show that the mm and cm variations are
correlated (Zhao et al. 2001a). Thus, the fact that the emission at
different frequencies is produced on different spatial scales (Melia et
al. 1992), suggests that this period should be induced by a single
process, which can cause correlated fluctuations across a broad range of
wavelengths. Second, the period is four orders of magnitude longer than
the dynamical time scale at the marginally stable orbit $r_{ms}$. So it
can either be produced on a much larger spatial scale, or it may be
associated with an intrinsic property of the supermassive black hole
itself, such as a nonzero spin. However, recent VLA observations have
indicated that the 2 cm emission is produced within a region no larger
than $140\ r_S$ (Bower et al. 2002), for which the corresponding dynamical
time scale is about one and a half days, which is much smaller than the
observed period. Higher frequency emission is expected to be produced
within still smaller regions, associated with even smaller time scales.

We may ask then, whether this modulation could be produced by a
corrugation wave in an accretion disk, which is used to account for the
quasi-periodic oscillations (QPOs) seen in low-mass X-ray binaries. These
waves have periods that are much longer than the corresponding dynamical
time scale (on a comparable spatial scale; Kota 1990), but the periods
depend on radius and thus may not be able to account for the first feature
described above. Moreover, the observed light curves show quite stable
periodic fluctuations, rather than the uncorrelated segments constituting
QPOs.  We thus conclude that the period must be associated with a single
process evolving on a small spatial scale (i.e., certainly less than $\sim
100\ r_S$ in size).

A schematic diagram of the assumed geometry for this system is shown in
Figure 1.  The oscillation frequency of a perturbation in the
perpendicular direction to a circular orbit in the equatorial plane of a
Kerr black hole has been derived by Kato (1990), and may be written in
the form 
$\nu_\theta^2=\nu_\phi^2\,(1\mp4\ a\ M^{1/2}r^{-3/2}+3\ a^2r^{-2})\;,$ 
where $G=c=1$, and 
$\nu_\phi=\pm M^{1/2}r^{-3/2}\left[2\pi\,(1\pm a\ 
M^{1/2}r^{-3/2})\right]^{-1}$
is the frequency of a circular orbit at radius $r$ measured by a static
observer at infinity (Bardeen et al. 1972). For small values of $a/M$,
the nodal precession frequency is then
$\nu_{p\theta}\equiv\nu_\phi-\nu_\theta\simeq a\ M\ r^{-3}/\pi\;.$
Because this frequency depends on $r$, the final configuration of a disk
that is misaligned with the black hole spin at large radii depends on
whether the viscous time scale $t_{vis}$ is shorter or longer than the
precession period ${\nu_{p\theta}}^{-1}$.  For a thin cold disk, the
differential Lense-Thirring precession will dominate the internal coupling
of the plasma at small radii and will therefore lead to the so-called
Bardeen-Petterson effect (Bardeen \& Petterson, 1975), in which the inner
region flattens toward the equatorial plane, producing a warped accretion
pattern.  As shown by Nelson \& Papaloizou (2000), however, thicker disks
with a mid-plane Mach number of $5$ or less can suppress warping
effects due to the coupling provided by pressure gradients in the gas. The
mid-plane Mach number in Sgr A* is $\sim 3$ (Melia, Liu, \& Coker 2001),
so it appears that the disk in this system precesses more or less as a
rigid body.

Since the surface density of the disk is roughly a constant (Melia, Liu,
\& Coker 2001), the total precessional torque acting on it is 
\begin{equation}
T=\sin{\xi}\int_{r_i}^{r_o} 2\pi\nu_{p\theta}\ v_k\ r\ 2\pi\ r\ 
\Sigma dr=8\pi\sin{\xi}\ a\ M (G\ M)^{1/2} \Sigma \left[1-(
r_i/r_o)^{1/2}\right]/r_i^{1/2},
\end{equation}
where $\xi$ is the angle between the axis of the disk and the black 
hole spin, $\Sigma$ is the surface density, $v_k$ is the Keplerian 
velocity and $r_i$ and $r_o$ are, respectively, the inner and outer
radii. The disk's total angular momentum is given by
$L=0.8\pi (G\ M)^{1/2} \Sigma\ r_o^{5/2}\left[1-(r_i/r_o)^{5/2}\right].$
Thus, the precession period of the disk is
\begin{equation}
P=2\pi\sin{\xi}(L/T) = {\pi r_o^{5/2}r_i^{1/2}\left[1 - 
(r_i/r_o)^{5/2}\right] \over 5\ a\ M \left[1-(r_i/r_o)^{1/2}\right]}.
\end{equation} 
In these calculations we have neglected the higher order general 
relativistic effects, which will introduce corrections of order 
$r_S/r$. The gas will fall into the black hole very quickly once 
it crosses $r_{ms}$. The inner boundary $r_i$ of the accretion disk may
therefore be taken to be $r_{ms}$, which is $\sim 3\ r_S$ for a black 
hole with small spin. Corrections introduced by the higher order 
general relativistic effects should then be no bigger than $\sim 30\%$.

The disk's outer radius, on the other hand, is not well constrained,
even though several observational lines of evidence point to a compact,
hot, magnetized, Keplerian region as the dominant radiator (Melia, Liu, 
\& Coker 2000; 2001; Liu \& Melia 2001). Earlier hydrodynamical
simulations, however, do suggest that the accretion flow in Sgr A* 
circularizes at a radius of $20-800\;r_S$ (Coker \& Melia 1997). In the 
following discussion, we will take $r_o=30\ r_S$ as a fiducial outer 
boundary. It is straightforward to see that an identification of this 
precession period with the detected radio modulation then implies a black 
hole spin
\begin{equation}
a/M=0.088\,(r_i/3\,r_S)^{1/2}(r_o/30\ r_S)^{5/2}\;{1 -
(10\ r_i/r_o)^{5/2} \over1-(10\ r_i/r_o)^{1/2}}\;.
\end{equation}  
(Note that in these units, $r_S=2\ M$.)

There are several questions that arise from this discussion. The first
concerns whether or not the accreted angular momentum vector can disrupt the
small inner disk as it precesses around the black hole's spin axis, since
$t_{vis}\ll P$ (Liu, \& Melia 2002). It should be emphasized, however, that
the viscous time scale is the time required for the gas to flow through the
disk, while the time required to realign the disk via accretion is determined
by the disk's total angular momentum vector and the angular momentum flux
through it. This depends primarily on whether or not angular momentum is
transported efficiently through the disk. The infalling gas carries angular
momentum inward. However, a magnetic stress applied to the flow by the accretor
can introduce an outward angular mementum flux that can effectively cancel 
much of the inward flux.  Recent work on processes below $r_{ms}$ has shown 
that the torque at the inner edge of the disk is not necessarily zero, and 
that angular momentum can therefore be extracted from the magnetized gas 
below $r_{ms}$ (see, e.g., Krolik 1999; Gammie, 1999).  Consequently, the 
angular momentum flux through the disk may be sufficiently small (perhaps 
even zero) that its effect on the inner structure may be
negligible. However, the disk can be disrupted by significant fluctuations 
in the angular momentum induced by variations in the large scale accretion 
flow.  As we alluded to earlier, this type of event is expected to
occur roughly every few hundred years (Coker \& Melia 1997).

Second, how does a precessing disk affect the cm emission, given that
earlier work has already established the implausibility of an accretion
flow producing this radiation via a {\sl thermal} synchrotron process (Liu
\& Melia 2001)?  One possibility is that the $\sim 1-20$ cm spectrum of
Sgr A* is produced by shock-accelerated non-thermal particles in the
region outside $\sim 30\,r_S$, where the infalling gas is circularizing.  
In that case, the radio modulation could be the result of optical depth
effects as the flattened distribution of particles precesses about the 
black hole's spin axis.  One should note, however, that due to the
weak internal coupling in the quasi-spherical region (Liu, \&
Melia 2001), the precessing disk should not affect this large 
scale accretion. A more interesting possibility is suggested by the fact that
the integrated $\sim 1-20$ cm luminosity of Sgr A* would be comparable to
the power extracted from its spin energy via a Blandford-Znajek type of
electromagnetic process if $a/M\approx 0.088$.  The 106-day modulation
would then presumably arise when the precessing disk periodically shadows
the non-thermal particles flooding the region surrounding the black hole
as they escape from their creation site near the event horizon.

For this mechanism to work, however, the vacuum must break down as $r$
approaches $r_S$, but one can see right away that if $a/M$ is indeed
as low as the value inferred from the radio luminosity of Sgr A*, this 
break down cannot happen with $\gamma$-rays produced via curvature 
radiation (Blandford \& Znajek 1977). Instead, the
prevalence of mm/sub-mm photons offers the possibility that inverse
Compton scatterings may produce a sufficient number of energetic
$\gamma$-rays that can materialize to form the required sea of
electron-positron pairs. To quantify this thought, we note that Sgr A*'s
sub-mm spectrum turns over at $\nu_{mm}\approx 236$ GHz with a flux
density of $2.8$ Jy (Falcke et al. 1998). The corresponding radiation
energy density near the event horizon is therefore $1.02$ erg cm$^{-3}$
(assuming a distance to the Galactic Center of $8.5$ kpc) which is larger
than the critical radiation energy density of $0.57$ ergs cm$^{-3}$
required to initiate the cascade of electron-positron pair creation (see
Eq. 2.9 of Blandford \& Znajek 1977). To break down the vacuum, this
photon number density must be matched to an adequately intense magnetic
field: 
$ B\geq 29\, (0.088\,M/a)(2.6\times 10^6 {\,M_\odot}/M)(236{\rm 
GHz}/\nu_{mm}) \;\hbox{gauss}\;.$
By comparison, the magnetic field in the accretion disk is inferred to be
about $20$ gauss (Melia, Liu, \& Coker 2001). However, Krolik (1999) has 
argued that below $r_{ms}$, the magnetic field energy density
will increase inward and should saturate to the local rest mass energy
density of the infalling gas. Gammie (1999) has built a specific model
where the magnetic field energy density increases significantly inside
$r_{ms}$. Based on these estimates, we would therefore expect the
magnetic field to be even larger than $29$ gauss near the event horizon.  
Under these conditions, the power extracted from the spinning black hole
would scale in the following fashion: 
$L\sim 1.1\times 10^{34}\, (B/29\,{\rm 
gauss})^2(a/0.088\,M)^2\,(M/2.6\times 
10^6 {\,M_\odot})^2\;\;\hbox{ergs s}^{-1}\;,$ 
a value that is based on the integrated $\sim 1-20$ cm radio emissivity of 
Sgr A*.

Unfortunately, the actual process by which the energy is transferred from
the black hole to the particles is not known with sufficient certainty
for us to predict with confidence the characteristics of their radiative
output, other than their maximum available power.  Nonetheless, by
assuming an equipartition between their energy density and that of the
magnetic field $B$, one can show that the angular size $\theta$ and the
value of $B$ for this non-thermal source may be inferred from its observed
turnover frequency $\nu_{mm}$ and the corresponding flux density $F_{mm}$
(Marscher 1983):
$\theta = \left[8\pi\times 10^{10}
\,n(\alpha)/b(\alpha)^2\,(2\alpha-1)\,(\gamma_1\ m_e\ c^2)^{2\alpha-1}
D_{\rm Gpc}\right]^{1/(4\alpha+15)} \nu_{mm}^{-1}
F_{mm}^{(2\alpha+7)/(4\alpha+15)}\;$ mas,
$B=10^{-5}\,b(\alpha)\,\left[8\pi\times 10^{10}
\,n(\alpha)/b(\alpha)^2\,(2\alpha-1)\,(\gamma_1\ m_e\ c^2)^{2\alpha-1}
D_{\rm Gpc}\right]^{4/(4\alpha+15)}\nu_{mm} 
F_{mm}^{-2/(4\alpha+15)}$ gauss,
where $\nu_{mm}$ and $F_{mm}$ are in units of GHz and Jy 
respectively, $\alpha$ is the spectral index of the power-law
synchrotron emission in the optically thin region, $n(\alpha)$ and
$b(\alpha)$ are parameters given in Marscher (1983),
$\gamma_1\, m_e\ c^2$ is the low-energy cutoff of the electron
distribution, and $D_{\rm Gpc}$ is the distance to the source in Gpc.  
The results are not very sensitive to the spectral index $\alpha$. For
Sgr A*, adopting an $\alpha$ of $1$, we find that the $22$ cm radio
emission ($\approx 0.53$ Jy) is produced within a region whose projected
size is $4.7$ mas, corresponding to $\approx 760\,r_S$.  Its magnetic
field intensity is $0.32$ gauss. At $1.3$ cm (with a flux of $\approx
1.1$ Jy), where the 106-day period was detected, the corresponding source
size is $65\,r_S$. We therefore conclude that a precessing disk
with a diameter of $\sim 60\,r_S$ can shadow a significant fraction of
the emitting region at $1.3$ cm, and thereby produce a noticeable
modulation at this wavelength, but not at $22$ cm.  This expected
behavior is consistent with what was observed, in that the amplitude of
the modulation was greatest at $1.3$ cm, dropping monotonically toward
longer wavelengths, and finally falling below statistically significant
levels beyond $6$ cm.

\section{Discussion and Conclusions}

If this picture is correct, there are several immediate consequences for
the nature of Sgr A*'s spectrum at mm/sub-mm and X-ray wavelengths.  If we
assume that the non-thermal emission from this process extends into the
sub-mm region with a spectral index of $0.2$ and it turns over at $236$
GHz, then the flux density at the turnover is $\approx 1.5$ Jy. The
calculated source size and magnetic field intensity at that frequency are
then $7.4\,r_S$ and $49$ gauss, respectively, which are also consistent
with the magnetic field inferred from our accretion model.  But the fact
that the {\sl observed} $236$ GHz flux density ($\approx 2.8$ Jy) is
larger than that produced by the non-thermal particles then points to the
need for a hot accretion disk and its contribution to the mm/sub-mm bump.

Thinking about the spectrum produced by the non-thermal particles in the
region beyond the observed break at $\nu_{mm}\approx 236$ GHz, we note
that extending the best fit X-ray spectral index (Baganoff et al. 2001) of
$1.2$ backward toward the mm/sub-mm region predicts a flux density of
$\approx 1.4$ Jy at $\nu_{mm}$, which is very close to the value one
obtains by extrapolating the non-thermal radio emission toward this
frequency (Liu \& Melia 2001). Although this may itself be a coincidence,
the convergence of Sgr A*'s low- and high-energy spectral components at
$\approx 236$ GHz suggests that this may therefore be the turnover
frequency that separates the optically thick and thin non-thermal emission
components in Sgr A*. The {\sl Chandra}-detected X-rays in the quiescent
state would then be due to optically-thin non-thermal synchrotron
processes. In this instance, however, $a/M$ would need to be somewhat
larger than $\approx 0.088$ in order to accommodate the implied
non-thermal particle luminosity of $2.9\times 10^{34}$ ergs s$^{-1}$.

There is clearly much to be done with this picture.  Not coincidentally,
the physics of accretion below the marginally stable orbit has attracted
some attention in recent years (Krolik 1999; Gammie 1999; Agol \& Krolik
2000). It appears that a nonzero stress at this radius can change the
properties of the accretion disk significantly, and strong dissipation
may be possible inside this orbit. Since the sharp increase in radial
velocity toward the event horizon leads to a significantly lower number
density, a continued equipartition between the particles and the magnetic
field may produce very high-energy particles via magnetic dissipation.  
Under some circumstances, this combination of low particle density and
magnetic reconnection may permit some particles to escape and form a
directed, relativistic outflow, a scenario that provides an alternative
to the trapped non-thermal particle model we have been describing here.
Further study of these various possibilities is warranted.

We note, finally, that the shadowing effect from the precessing disk
should be even more pronounced at $43$ GHz than at $22$ GHz, based on our
estimate for the relative sizes of the emitting regions, suggesting that
an observation of a 106-day period at this higher frequency should reveal
an even stronger modulation than has been seen thus far at the longer
wavelengths.

We are very grateful to the anonymous referee, whose comments have led to
a significant improvement of the manuscript. This research was partially
supported by NASA under grants NAG5-8239 and NAG5-9205, and has made use
of NASA's Astrophysics Data System Abstract Service.  FM is grateful to
the University of Melbourne for its support (through a Miegunyah
Fellowship).

\vfill
\newpage
\begin{figure}[!htb]
\begin{center}
\includegraphics[width=0.7\textwidth]{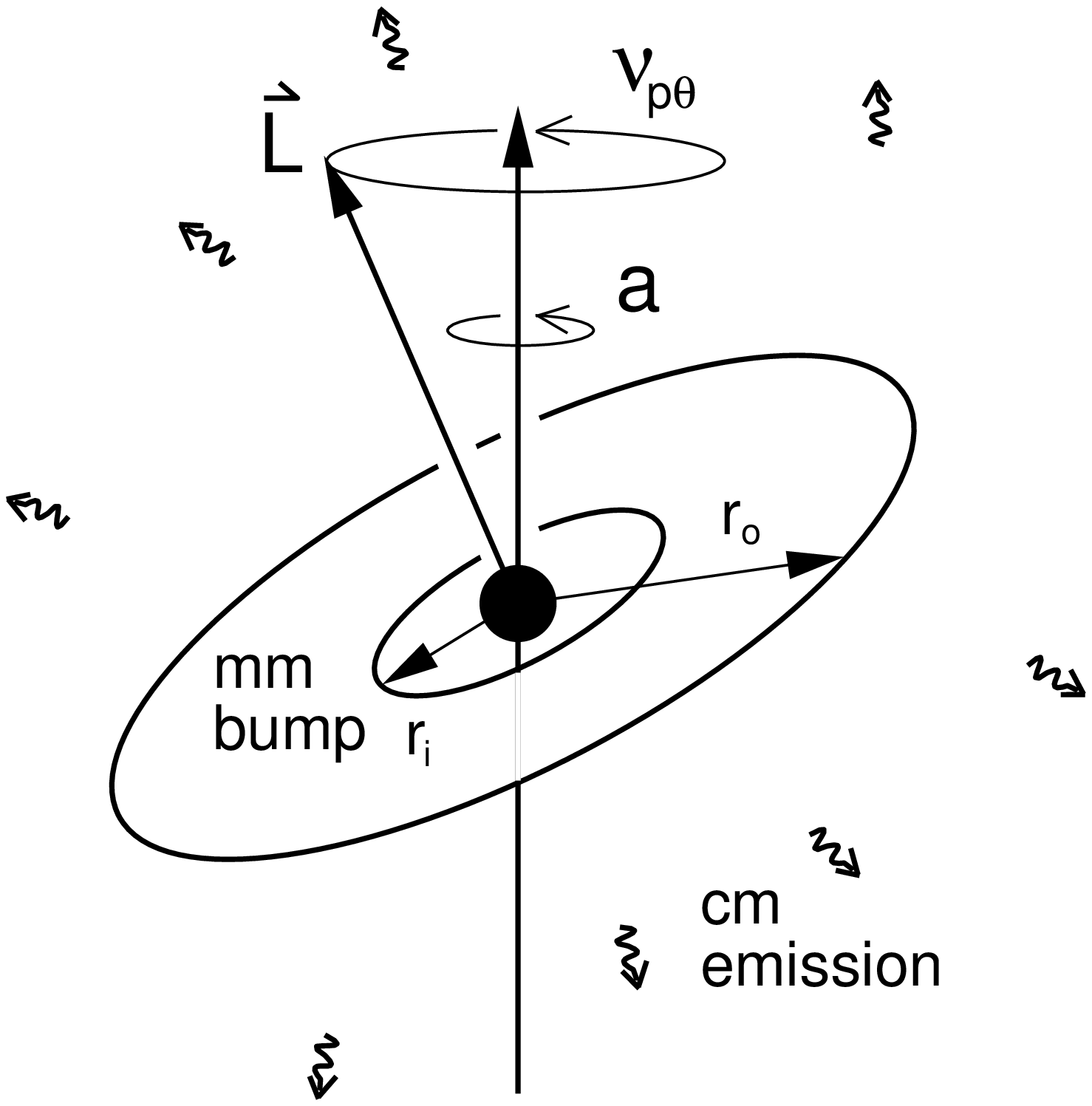}
\vskip 1.3in
\caption{Schematic diagram showing the pertinent
characteristics of a system in which the spin-induced precession
of a compact disk accounts for the observed 106-day cycle in the
$\lambda>1$ cm emission from Sgr A*.  The mm/sub-mm bump in Sgr
A*'s spectrum is thought to originate from the hot, magnetized
Keplerian structure, whereas the $\sim 1-20$ cm emission is
probably due to the synchrotron emission of non-thermal particles, 
possibly energized by a Blandford-Znajek process near the 
event horizon. Whether or not these particles escape from
the system and form a directed outflow may depend on the strength
of the magnetic field and its configuration.  In either case,
the 106-day modulation may be due to the periodic shadowing of
the non-thermal emitting region by the precessing disk. Here, $\vec L$
is the angular momentum vector of the disk, $a$ is the spin parameter
for the black hole, and $\nu_{p\theta}$ is the precession frequency
of the disk.}
\label{fig:schematic}
\end{center}
\end{figure}

\end{document}